\documentclass[prb, twocolumn, showpacs, preprintnumbers, amsmath, amssymb]{revtex4}

\pdfoutput=1
\usepackage{graphicx, color, graphpap}% Include figure files
\usepackage{dcolumn}% Align table columns on decimal point
\usepackage{bm}% bold math

\begin{document}

\title{Helicity asymmetry of optically-pumped NMR spectra in GaAs}

\author{Patrick J. Coles}
\affiliation{Department of Chemical Engineering, University of California, Berkeley; Berkeley, CA 94720}

\date{\today}

\begin{abstract}

The origin of $\sigma\pm$ asymmetries in the optically-pumped NMR signal and hyperfine shift in GaAs is derived analytically and tested experimentally. The ratio of the optically-pumped to the equilibrium electron polarizations is a key parameter in determining both asymmetries. Variations in asymmetry with photon energy and laser power reflect variations in the local temperature and the electron spin polarization, and these two quantities are extracted from the asymmetry through a simple methodology. Other contributions to the asymmetry are considered.

\end{abstract}

\maketitle

\begin{figure}
\center
\includegraphics[width=8cm]{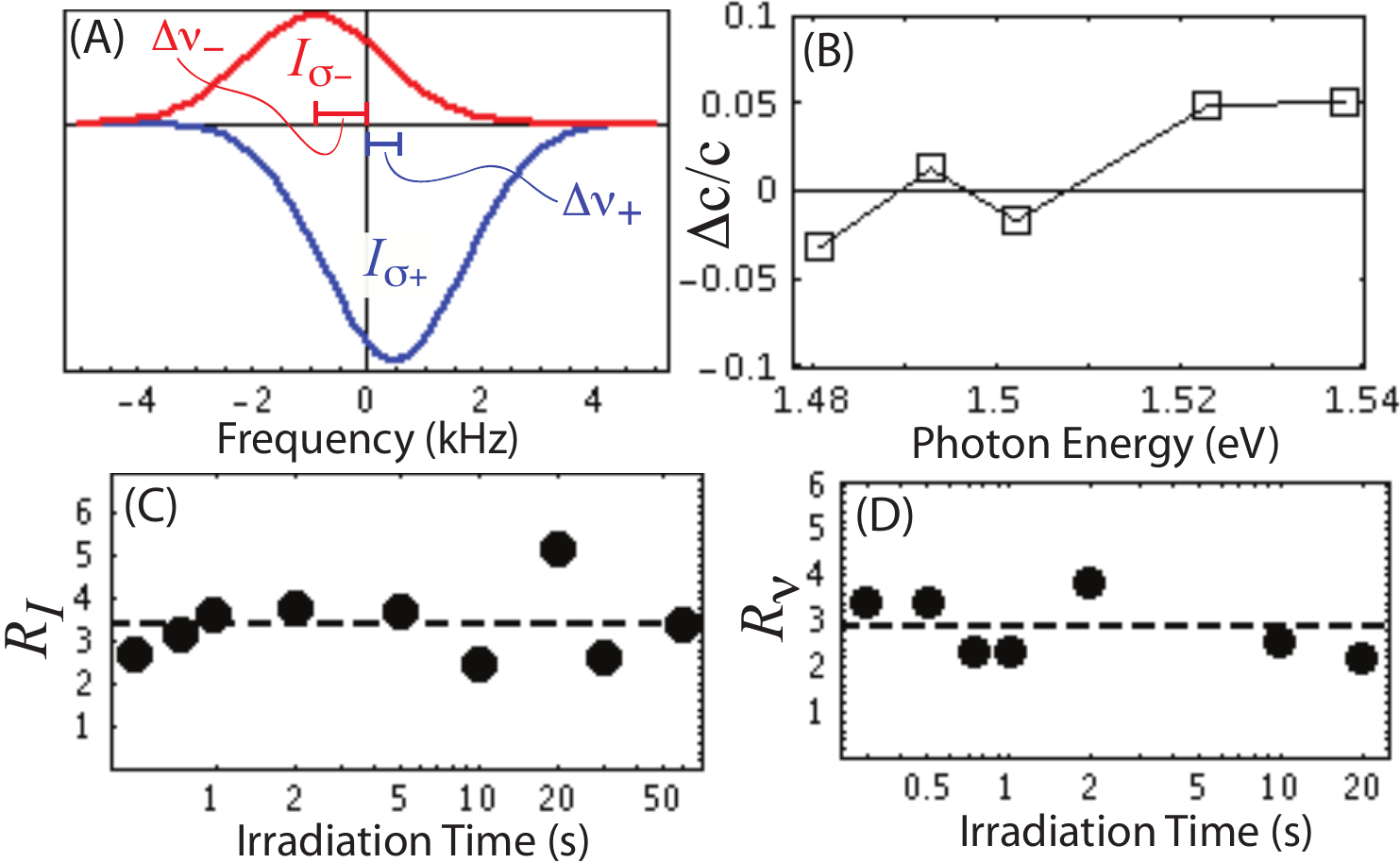}
\caption{(A) Gaussian fits to $\sigma\pm$ $^{71}$Ga OPNMR spectra for $t$=0.5 s, $P$=123 mW, $E$=1.509 eV, $B_0$=9.4 T, $T$=9.5 K. (B) Photoconductivity asymmetry, (C) $R_I$, and (D) $R_{\nu}$ under these conditions. Dashed lines indicate average values.}
\label{S0SsRiRv}
\end{figure}

Optical polarization of nuclear spins in semiconductors has sparked high sensitivity NMR studies of the quantum Hall regime,\cite{OPNMRQWs} spatial mappings of lattice strain,\cite{StrainOPNMR} and proposals for biological polymers.\cite{TyckoBio} In the optically-enhanced or optically-pumped NMR (OPNMR) spectrum, $\sigma\pm$ asymmetry has been noted in both the hyperfine-induced frequency shift\cite{HyperfineShift, OPNMRwashu, ColesReimer2007} and the NMR signal magnitude,\cite{TyckoStrayField, OPNMRwashu, MuiEtAl, ParavastuReimerPRB2004} as shown in Fig. \ref{S0SsRiRv}A. Variations in the asymmetry with photon energy\cite{OPNMRwashu, MuiEtAl, ParavastuReimerPRB2004} and laser power (shown herein) have not yet been explained, even for the most well-studied semiconductor, GaAs. Varying values of $\tau/T_{1e}$, where $\tau$ and $T_{1e}$ are the electron recombination and spin-relaxation times, were recently invoked to explain variations in signal magnitude asymmetry in GaAs.\cite{MuiEtAl} It is shown herein that the standard OPNMR model for GaAs predicts that $\tau/T_{1e}$ only affects the hyperfine shift asymmetry. 

It is often assumed that the initially-excited electron spin polarization ($\langle S_z\rangle_0\equiv S_0$) with $\sigma\pm$ light in bulk semiconductors is constant ($\pm$50\% for cubic crystals) for photon energies $E$ in the range $E_g < E < E_g+\Delta$ (where $E_g$ is the band-gap energy and $\Delta$ is the spin-orbit splitting),\cite{OpticalOrientation} and there is very little literature on optical electron spin orientation with $E<E_g$. By dropping the assumption that $S_0$ is a constant, we find an explanation for the varying asymmetry of OPNMR spectra. It furthermore follows that this asymmetry is related so simply to electron spin parameters that it can be used to measure the electron spin polarization. Thus, this article outlines a ``spinometry" method that, because the hole spin-nuclear spin interaction is very weak, probes the \textit{electron} spin polarization under optical pumping.

It was shown previously that the OPNMR signal in GaAs strongly correlates with the photoconductivity, and that the excitation spectrum of OPNMR signals can be largely understood from a simple picture in which optical absorption generates free electrons which then bind to shallow donors, analogous to the way that a gas adsorbs to a solid surface with a fixed number of sites.\cite{ColesReimer2007} Rapid spin exchange\cite{SpinExchange} maintains the steady-state polarizations of the free and donor-bound electron reservoirs to be equal and thus given by a single equation:\cite{ColesReimer2007, BowersGaAsSSNMR, OPspintemp, SsEquationReference}
\begin{equation}
S_s=\frac{S_0+S_{eq} \tau/T_{1e}}{1+\tau/T_{1e}},
\label{EqnSs}
\end{equation}
where $S_{eq}$ is the Boltzmann electron polarization. Once bound to shallow donors, the electrons experience a strong hyperfine interaction and can undergo mutual spin flips with nearby nuclear spins. The dimensionless nuclear polarization, $C=\langle I_z \rangle/I_s$, near the shallow donors then evolves over space and time according to a generation-diffusion-loss equation:\cite{ShallowDonors1982, BowersGaAsSSNMR, MagneticFieldDependence, ColesReimer2007}
\begin{equation}
\frac{\partial C}{\partial t}=D \nabla^2 C+\frac{1}{T_{1H}(\vec{x})}(1-C)-\frac{1}{T_{1L}(\vec{x})}C
\label{EqnCevolution}
\end{equation}
where $I_s$ is the theoretical maximum nuclear polarization achievable through cross relaxation, $D$ is the nuclear spin diffusivity, $T_{1H}$ is the hyperfine cross-relaxation time, $T_{1L}$ is the nuclear spin-lattice relaxation time by all other mechanisms, and the Boltzmann nuclear polarization is neglected. During this microscopic evolution, the bulk NMR signal $I(t)$ grows in proportion to the total nuclear $z$-angular momentum, found by integrating $\langle I_z \rangle$ over all space: $I(t)$=$\int \langle I_z \rangle(\vec{x},t) dV$=$\int I_s C(\vec{x},t) dV$. In the high lattice temperature approximation, $I_s$ is given by:\cite{Abragam} $I_s$=$\frac{I(I+1)}{S(S+1)} \kappa (S_s - S_{eq})$=$\frac{I(I+1)}{S(S+1)} \kappa \frac{S_0 - S_{eq}}{1+\tau/T_{1e}}$. Here, $I$ and $S$ are the nuclear and electron spin quantum numbers and $\kappa$=$(w_0-w_2)/(w_2+w_0+2w_1)$, where $w_0$, $w_2$, and $w_1$ are the transition probabilities per time for electron-nuclear flip-flops, flip-flips, and independent nuclear flips. Thus, the NMR signal follows a complicated equation:
\begin{equation}
I(t)= (S_0 - S_{eq})\int \frac{I(I+1)}{S(S+1)} \frac{\kappa C(\vec{x},t)}{1+\tau/T_{1e}} dV.
\label{EqnNMRsignal}
\end{equation}

This expression can be reduced to something simple through a transformation. Consider that $S_0$ varies sinusoidally with the angle $\theta$ of the quarter-wave retarder, through which the laser light passes, according to $S_0 = \Lambda \sin 2\theta$. ($\Lambda$ is treated as an experimentally determined quantity in this paper.) Ignoring any dependence of $S_{eq}$, $\tau$, $T_{1e}$, and $C(\vec{x},t)$ on the light helicity, Eq. (\ref{EqnNMRsignal}) gives $A\equiv(I_{\sigma +}-I_{\sigma -})$=$k(t)\Lambda$ and $B\equiv(I_{\sigma +}+I_{\sigma -})$=$k(t)S_{eq}$, where $I_{\sigma \pm}$ denotes the NMR signal induced by $\sigma\pm$ light. All the time dependence and all the nuclear spin parameters are contained in $k(t)$$\equiv$$ -2\int \frac{I(I+1)}{S(S+1)} \frac{\kappa CdV}{1+\tau/T_{1e}}$. So the \textit{ratio} of the two equations cancels out everything but the electron spin polarizations:
\begin{equation}
R_I \equiv \frac{I_{\sigma +}-I_{\sigma -}}{I_{\sigma +}+I_{\sigma -}}= \frac{\Lambda}{S_{eq}}.
\label{EqnRatio}
\end{equation}
This simple relationship provides an interpretation for the asymmetry of OPNMR signals, and forms the basis for extracting $S_0$ from the signal asymmetry:
\begin{equation}
S_0=S_{eq} R_I \sin 2\theta.
\label{EqnS0Extraction}
\end{equation}

A similar procedure can be followed for the hyperfine shift of the OPNMR line. This shift is given by: $\Delta\nu(t)$=$\Delta\nu_m\int S_sFe^{-r/a_0}\langle I_z \rangle dV/\int \langle I_z \rangle dV$, where $\Delta\nu_m$ is a constant specific to the nuclear isotope, $r$ is the distance to the nearest shallow donor, $F$ is the occupation probability of that donor, and $a_0$ is its Bohr radius.\cite{HyperfineShift, ColesReimer2007} Using Eq. (\ref{EqnSs}) and assuming that $\tau/T_{1e}$ (and thus, $S_s$) is spatially uniform:
\begin{equation}
\Delta\nu(t)= \Delta\nu_m(S_0+\frac{\tau}{T_{1e}}S_{eq})\frac{\int Fe^{-r/a_0}CdV}{(1+\tau/T_{1e})\int CdV}.
\label{EqnHyperfineShift}
\end{equation}
Denoting the hyperfine shift with $\sigma_{\pm}$ light as $\Delta\nu_{\pm}$ and making the same approximations as those made in deriving $R_I$, we obtain $(\Delta\nu_--\Delta\nu_+)$=$m(t)\Lambda$ and $(\Delta\nu_-+\Delta\nu_+)$=$m(t)\frac{\tau}{T_{1e}}S_{eq}$. All the time dependence is contained in $m(t)$$\equiv$$2\Delta\nu_m \frac{\int Fe^{-r/a_0}CdV}{(1+\tau/T_{1e})\int CdV}$. Again, taking the ratio of the two equations gives an interpretation for the asymmetry of hyperfine shifts:
\begin{equation}
R_{\nu}\equiv\frac{\Delta\nu_--\Delta\nu_+}{\Delta\nu_-+\Delta\nu_+}=\frac{\Lambda}{\frac{\tau}{T_{1e}}S_{eq}}.
\label{EqnRatioNu}
\end{equation}
Just like Eq. (\ref{EqnRatio}), this is a simple relationship between measurable \textit{NMR} quantities and time-independent \textit{electronic} parameters. Equations (\ref{EqnSs}), (\ref{EqnRatio}), and (\ref{EqnRatioNu}) may be combined to solve for $S_s$:
\begin{equation}
S_s=S_{eq} \frac{R_I \sin 2\theta+R_I/R_{\nu}}{1+R_I/R_{\nu}}.
\label{EqnSsExtraction}
\end{equation}

The above derivation shows that $\Lambda/S_{eq}$ is a key parameter in determining the asymmetry of OPNMR spectra. Other contributions to the asymmetry may result from effects such as inhomogeneous Knight fields,\cite{InhomoHyperfine} large Overhauser nuclear fields, and spin-dependent recombination\cite{SDR,Brunetti} (SDR).\footnote{Also note that a non-negligible Boltzmann nuclear polarization in Eq. (\ref{EqnCevolution}) could affect the asymmetry at high magnetic fields and conditions that favor short $T_{1L}$.} Respectively, these effects would impart a helicity dependence to $D$, $S_{eq}$ (through the magnetic field), and $\tau$. The helicity dependence of $\tau$ can be measured by the helicity dependence of the photoconductivity. Figure \ref{S0SsRiRv}B plots the photoconductivity asymmetry: $\Delta c/c\equiv(c_{\sigma+}-c_{\sigma-})/(c_{\sigma+}+c_{\sigma-})$, which is attributed to SDR, observed for laser power $P\approx$200 mW at the $B_0$, $T$, and $E$-range under which our OPNMR measurements were performed. The data indicate that SDR is present in semi-insulating (SI) GaAs and is larger for $E>E_g$, but overall it appears to be small: $\Delta c/c$ was $\lesssim 5\%$ over this photon energy range. So the helicity dependence of $\tau$ appears not to be the dominant contribution to the OPNMR asymmetry. SDR may also alter the form of Eq. (\ref{EqnSs}), although this would go beyond the standard model for OPNMR in GaAs,\cite{ColesReimer2007, HyperfineShift, BowersGaAsSSNMR, MagneticFieldDependence, ShallowDonors1982, OPNMRwashu} and should be addressed in future work. The other two contributions to the OPNMR asymmetry are irradiation-time dependent: the hyperfine-blocked nuclear spin diffusion affects signal-growth kinetics at short irradiation times, whereas the Overhauser nuclear field increases with irradiation time. Figures \ref{S0SsRiRv}C and \ref{S0SsRiRv}D plot the irradiation-time dependence of the signal and shift asymmetries, respectively. One can see that neither asymmetry exhibits a trend or correlation with time, in agreement with Eqs. (\ref{EqnRatio}) and (\ref{EqnRatioNu}), and suggestive that inhomogeneous Knight fields and Overhauser nuclear fields are not the dominant contributions to the asymmetry. Thus, data in this article are interpreted with Eqs. (\ref{EqnRatio}) and (\ref{EqnRatioNu}). In general, the presence of extra helicity-dependent quantities (besides $S_0$) could be diagnosed by the $\theta$-dependence of OPNMR signal deviating from the expression $(\alpha+\beta \sin 2\theta)$, but this dependence was followed very well in GaAs.\cite{BowersGaAsSSNMR}

With this interpretation, Eqs. (\ref{EqnS0Extraction}) and (\ref{EqnSsExtraction}) imply that the values of $S_0$ and $S_s$ may be extracted from the helicity asymmetry of the OPNMR spectrum. This requires that $S_{eq}$=$\frac{1}{2} \mbox{tanh} \frac{-g^*\mu_B B_0}{2 k_b T}$ be known, which essentially means knowing $T$ since the effective $g$-factor $g^*$ and the applied magnetic field $B_0$ are known.\footnote{This method also requires that any dark NMR signals be phase-cycled away from the OPNMR spectrum, say, with a spectrometer-controlled laser shutter.} The percent error in $T$ propagates to about the same percent error in $S_0$ and $S_s$, which is problematic during irradiation because local heating causes $T$ to deviate from a commercial cryostat's thermistor reading, $T_0$. Thus, we discuss below a laser heating model and a procedure to extract all three parameters: $\Lambda$, $T$, and $S_s$, from the OPNMR spectrum.

For an irradiated volume with heat capacity $C_p$ and surface area $2 \sigma$, where $\sigma$ is the laser spot area, absorbing a net power of $fP$ and losing heat to the surroundings (temperature $T_0$), an energy balance gives:
\begin{equation}
C_p\frac{dT}{dt}=fP-2h\sigma(T-T_0).
\end{equation}
The steady-state temperature $T_s$ grows linearly with $P$:
\begin{equation}
T_s = T_0 +f P/(2 h \sigma).
\end{equation}
This model further suggests that $T$ rises extremely fast after laser exposure due to the low $C_p$ at low $T$: the initial rise rate being $f P/C_p \sim$1 K/10 $\mu$s at 5 K and $P$=200 mW. Steady-state is reached typically within the first millisecond, which is very rapid compared to optical nuclear polarization, so $T$ can be assumed constant during the nuclear polarization process. 

\begin{figure}
\center
\includegraphics[width=8cm]{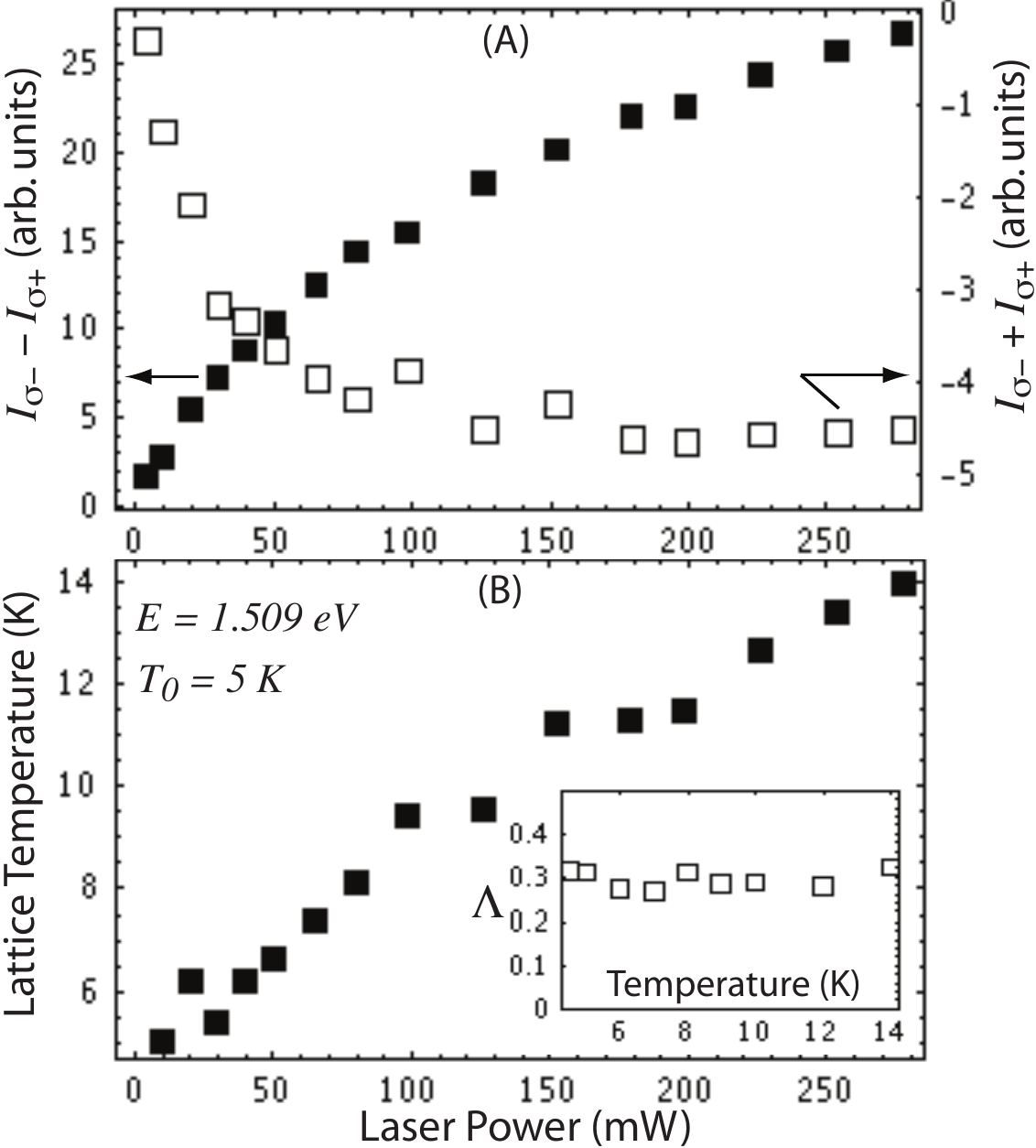}
\caption{(A) Power dependence of $A$ and $B$, defined in text, for $E$=1.509 eV and $T_0$=5 K. (B) Extracted $T$ versus power. Inset: Extracted $\Lambda$ over this temperature range.}
\label{LatticeTempVsPower}
\end{figure}

A procedure for extracting the parameters is:
(1) Measure $R_I$ at very low $P$ where $T\approx T_0$ and calculate $\Lambda=\frac{-g^*\mu_B B_o}{4k_bT_0}[R_I]_{P \rightarrow 0}$. In other words, extract $\Lambda$ from the y-intercept of the $R_I \hspace{2 pt} \mbox{vs.} \hspace{2 pt} P$ plot.
(2) Go to the $P$ of interest and extract $T$ using $S_{eq}=\Lambda/R_I$.
(3) Extract $S_s$ at this $P$ from Eq. (\ref{EqnSsExtraction}). Thus, the OPNMR signal serves as a spinometer \textit{and} a thermometer. This thermometry method requires no sample preparation and is an \textit{in situ} measurement of the irradiated volume's temperature. Figure \ref{LatticeTempVsPower}B shows a thermometry curve obtained with this method, the linear dependence on $P$ being consistent with the model's prediction. Also obtained with this method, the inset shows that $\Lambda$ at 1.509 eV is $\sim0.3$ and is practically $T$-independent from 4.5 K-14 K.

Laser heating is major concern in OPNMR. In previous studies of the $P$-dependence,\cite{OPsilicon, CdTe, TyckoBio, ColesReimer2007, OPNMRQWs} OPNMR signals from a single light polarization were  plotted vs. $P$, which had the disadvantage that the signal depended on $S_{eq}$, and thus was vulnerable to laser heating effects. The quantity $A$ is independent of $S_{eq}$, more robust to laser heating, easier to model, and better for studying nonlinear effects like $2^{nd}$ order carrier recombination and shallow-donor filling.\cite{ColesReimer2007} Figure \ref{LatticeTempVsPower}A plots $A$ and $B$ vs. $P$. While $B$ saturates at some $P$ value, $A$ keeps on growing. Previously, saturation of OPNMR signals with linear polarized light at high $P$ was attributed to complete filling of shallow donors,\cite{ColesReimer2007} but the data in Fig. \ref{LatticeTempVsPower} show that it was actually due to laser heating, and we can no longer say that we obtained proof of shallow-donor filling from the $P$-dependence of OPNMR. (This does not eliminate the possibility that shallow-donor filling is important at these powers.) So this thermometry method has had immediate impact on our understanding of OPNMR data.

\begin{figure}
\center
\includegraphics[width=8cm]{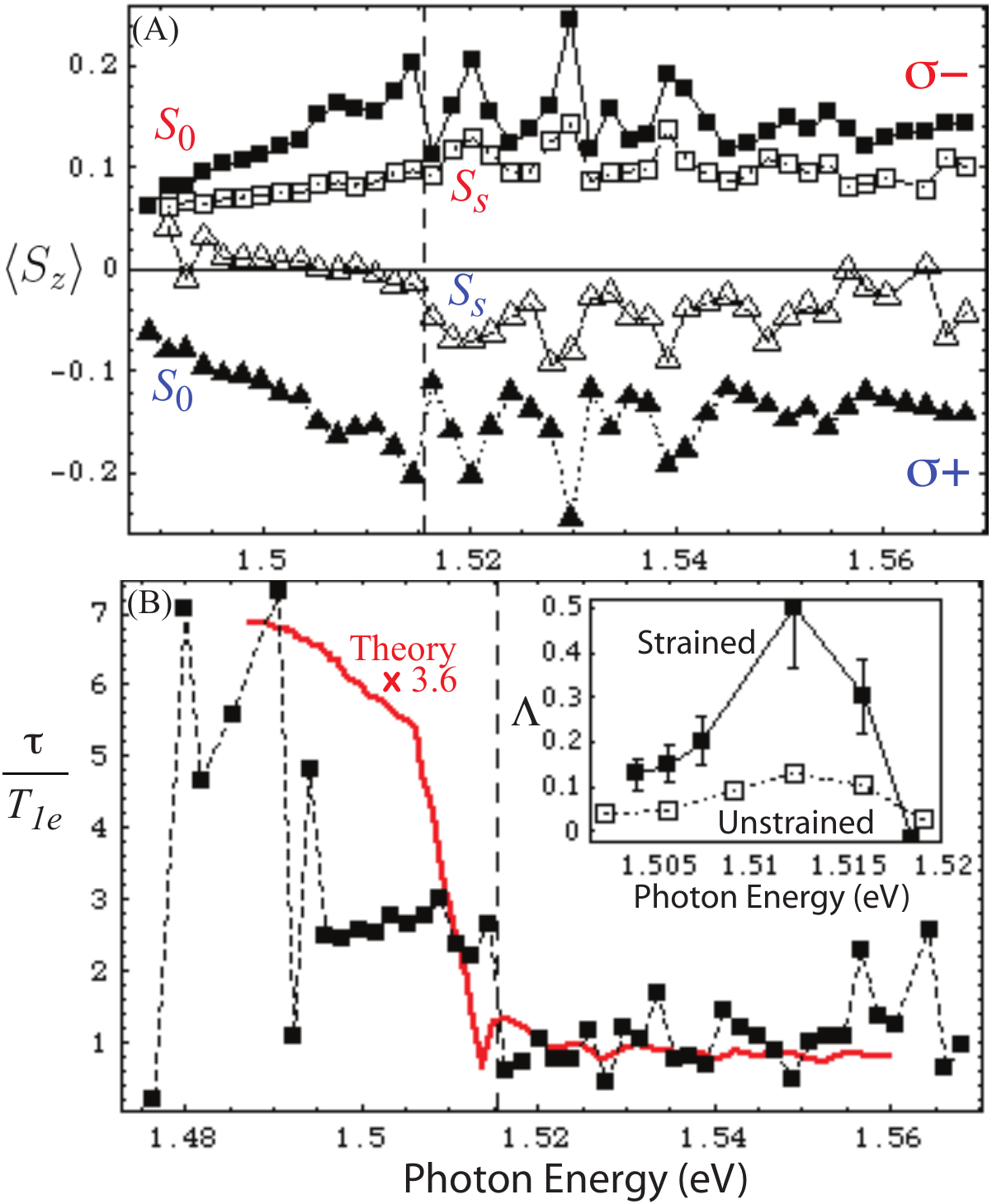}
\caption{(A) Excitation spectrum of $\langle S_z \rangle$ in SI GaAs from $^{71}$Ga data.\cite{OPNMRwashu} $S_s$ values (open) are bounded by $S_0$ values (filled), shown for $\sigma-$ (squares) and $\sigma+$ (triangles) light. (B) Spectrum of $\tau/T_{1e}$ from the same data. Overlaid is a theoretical prediction made by our previous model for OPNMR.\cite{ColesReimer2007} Unphysical (negative $\tau/T_{1e}$) data points were removed and dashed lines indicate the free exciton energy. Inset: Effect of strain on $\Lambda$ extracted from $^{71}$Ga data.\cite{ParavastuReimerPRB2004, OPspintemp}}
\label{SzEnergyDependence}
\end{figure}

$S_0$ and $S_s$ were also measured for SI GaAs with the above method. The NMR lineshapes were Gaussian for all $t$, so the hyperfine shift extracted from a Gaussian fit was equivalent to the first moment as defined above. For $E$=1.509 eV, $T_0$=5K, and $P$=123 mW, the calculated parameters are: $T$=9.5K, $\tau/T_{1e}$=$R_I/R_{\nu}$=1.2, $\Lambda$=0.294$\pm$0.03, $S_{s,\sigma-}$=0.184$\pm$0.03, and $S_{s,\sigma+}$=$-$0.088$\pm$0.02. The extracted $\Lambda$ is slightly greater than the theoretically predicted value of 0.25.

It has been derived analytically and demonstrated experimentally that the asymmetry of optically pumped NMR spectra in GaAs is a measure of $T$ and $\langle S_z\rangle$. This method is now applied to literature data to map out the $E$-dependence of $\langle S_z\rangle$.\footnote{It is unclear whether temperatures in the literature are corrected for laser heating. Thus, uncertainties on spin polarizations extracted from the literature cannot be provided. Not accounting for laser heating would overestimate $\langle S_z\rangle$.} Figure \ref{SzEnergyDependence}A shows the spectrum of $S_0$ and $S_s$ excited by $\sigma\pm$ light, for SI GaAs at $B_0$=4.7 T, $T$=6 K, $P$=2.5 W/cm$^2$.\cite{OPNMRwashu} There are several peaks and valleys, indicating that the $E$-dependence of $\langle S_z\rangle$ should be taken into account when modeling OPNMR $\sigma\pm$ data. Similar oscillations have been observed in GaAs quantum wells,\cite{OOQWs} so the strong magnetic field may provide the analogous confinement potential that breaks the degeneracy between heavy-hole and light-hole excitations and gives rise to oscillations in optical orientation.

$S_0$ gradually increases over a 30 meV region just below the band edge, a feature that was reproduced in all 8 OPNMR data sets that we inspected, including those for different nuclear isotopes and magnetic fields. A reasonable explanation is an increasing proportion of shallow defect-to-band relative to deep defect-to-band absorption as $E$ approaches $E_g$. Only the former transitions have the same selection rules (thus the same spin-selectivity) as band-to-band transitions. Shallow acceptors in GaAs sit $\sim$26 meV above the valence band,\cite{CarbonAcceptorPL} so onset of band-to-band-like optical electron spin orientation likely coincides with onset of shallow acceptor absorption. The lower value of $\Lambda/S_{eq}$ at low $E$ shown here may explain the observed inversion\cite{MuiEtAl} of $\sigma-$ OPNMR signal at low $E$.

Fig. \ref{SzEnergyDependence}B plots the ratio of the two asymmetries $R_I/R_{\nu}=\tau/T_{1e}$, which decreases by $\sim$1 order of magnitude as $E$ increases from below to above $E_g$. Overlaid is a theoretical spectrum ($\times$3.6 for comparison) predicted by our previous OPNMR model without adjusting any fitting parameters.\cite{ColesReimer2007} It accounts only for the variation of $\tau$ (assumes constant $T_{1e}$) and likewise predicts the decrease by 1 order of magnitude. So the main feature in the $E$-dependence of $\tau/T_{1e}$ may be explained by the $E$-dependence of $\tau$: the shorter penetration depth at higher $E$ confines the electron-hole gas to a smaller volume, increasing the recombination rate. \cite{ColesReimer2007} The lower $\tau/T_{1e}$ at higher $E$ pulls $S_s$ closer to $S_0$, as seen in Fig. \ref{SzEnergyDependence}A. This implies that super-gap irradiation leads to a better retention of the injected polarization in the steady state.

The asymmetry of literature $^{71}$Ga OPNMR signals\cite{ParavastuReimerPRB2004, OPspintemp} was different for strained and unstrained GaAs, as shown in the inset of Fig. \ref{SzEnergyDependence}B. The asymmetry difference is consistent with the strain enhancing the initially-excited electron spin polarization by about a factor of 2, which is precisely what is predicted for a compressive strain perpendicular to the magnetic field.\cite{OpticalOrientation} This enhancement of optical electron spin orientation with strain was not considered in previous OPNMR data interpretation.\cite{OPspintemp}

In conclusion, an interpretation was given for the asymmetry of OPNMR signals and hyperfine shifts in GaAs in terms of the local temperature and the electron spin polarization. This provided a methodology for extracting the latter two quantities from OPNMR asymmetries. This method may be used to investigate effects of quantum confinement, strain, and applied magnetic field on optical pumping, to probe spin-selective excitation of defects, and to study electron recombination and spin-lattice relaxation. The author thanks Jeff Reimer for helpful suggestions and support, and acknowledges support from the NSF under project ECS-0608763. 

\bibliographystyle{h-physrev3.bst}
\bibliography{BriefReport}

\begin{thebibliography}{10}

%\bibitem{PRLimagingSpinFlow}
%S.A. Crooker and D.L. Smith,
%Phys. Rev. Lett. \textbf{94}, 236601 (2005).

%\bibitem{SpinTransportwithOO}
%T. Taniyama, G. Wastlbauer, A. Ionescu, M. Tselepi, and J.A.C. Bland,
%Phys. Rev. B \textbf{68}, 134430 (2003).

%\bibitem{VCSEL}
%S. Hovel, N. Gerhardt, M. Hofmann, J. Yang, D. Reuter, and A. Wieck,
%Elect. Lett. \textbf{41}(5), (2005).

%\bibitem{VCSEL}
%S. Hovel, N.C. Gerhardt, C. Brenner, M.R. Hofmann, F.Y. Lo, D. Reuter, A.D. %Wieck, E. Schuster, and W. Keune,
%Phys. Stat. Sol. (A) \textbf{204}, 500 (2007).

%\bibitem{ResonantSpinAmplification}
%J.M. Kikkawa and D.D. Awschalom, Phys. Rev. Lett. \textbf{80}, 4313 %(1998).

%\bibitem{ElectronSpinDiffusion}
%S.G. Carter, Z. Chen, and S.T. Cundiff, Phys. Rev. Lett. \textbf{97}, %136602 (2006).

\bibitem{OPNMRQWs}
S.E. Barrett, R. Tycko, L.N. Pfeiffer, and K.W. West, Phys. Rev. Lett. \textbf{72}, 1368 (1994).

%\bibitem{OPNMRquantumHall}
%A.E. Dementyev, P. Khandelwal, N.N. Kuzma, S.E. Barrett, L.N. Pfeiffer, and K.W. %West, Sol. State Comm. \textbf{119}, 217 (2001).

\bibitem{StrainOPNMR}
M. Eickhoff, B. Lenzmann, D. Suter, S.E. Hayes, and A.D. Wieck, Phys. Rev. B \textbf{67}, 085308 (2003).

\bibitem{TyckoBio}
R.~Tycko, Sol. State Nuc. Mag. Res. \textbf{11}, 1 (1998).

\bibitem{HyperfineShift}
K. Ramaswamy, S. Mui, and S.E. Hayes, Phys. Rev. B \textbf{74}, 153201 (2006).

\bibitem{ColesReimer2007}
P. Coles and J. Reimer, Phys. Rev. B \textbf{76}, 174440 (2007).

\bibitem{OPNMRwashu}
S. Mui, K. Ramaswamy, and S.E. Hayes, Phys. Rev. B \textbf{75}, 195207 (2007).

\bibitem{TyckoStrayField}
C.A. Michal and R. Tycko, Phys. Rev. B \textbf{60}, 8672 (1999).

\bibitem{MuiEtAl}
S. Mui, K. Ramaswamy, and S.E. Hayes, J. Chem. Phys. \textbf{128}, 052303 (2008).

\bibitem{ParavastuReimerPRB2004}
A.K. Paravastu, S.E. Hayes, B.E. Schwickert, L.N. Dinh, M. Balooch, and J.A. Reimer, Phys. Rev. B \textbf{69}, 075203 (2004).

\bibitem{OpticalOrientation}
\textit{Optical Orientation}, edited by F. Meier and B.P. Zakharchenya
(Elsevier, Amsterdam, 1984).

%\bibitem{TheoryLambdaSpectrum}
%F. Nastos, J. Rioux, M. Strimas-Mackey, B.S. Mendoza, and J.E. Sipe, Phys. %Rev. B \textbf{76}, 205113 (2007).

\bibitem{SpinExchange}
D.~Paget, Phys. Rev. B \textbf{24}, 3776, (1981).

%\bibitem{KerrRotation}
%A.V. Kimel, A.A. Tsvetkov, A. Kirilyuk, Th. Rasing, and V.N. Gridnev
%Phys. Rev. B \textbf{63}, 235201 (2001).

%\bibitem{Parsons}
%R.R. Parsons, Phys. Rev. Lett. \textbf{23}, 1152 (1969).

%
%\bibitem{Bagraev}
%N.T. Bagraev, L.S. Vlasenko
%\newblock {\em JETP}, 28:527, 1978.
%
%\bibitem{Lampel}
%G. Lampel
%\newblock {\em Phys. Rev. Lett.}, 20:491, 1968.
%
%\bibitem{SiliconSpinInjection}
%I. Zutic, J. Fabian, and S.C. Erwin
%\newblock {\em Phys. Rev. Lett.}, \textbf{97}, 026602, 2006.

\bibitem{BowersGaAsSSNMR}
C.R. Bowers, Sol. State Nuc. Mag. Res. \textbf{11}, 11 (1998).

\bibitem{SsEquationReference}
C. Weisbuch and C. Hermann, Phys. Rev. B \textbf{15}, 816 (1977).

\bibitem{OPspintemp}
A.K. Paravastu and J.A. Reimer, Phys. Rev. B \textbf{71}, 045215 (2005).

\bibitem{ShallowDonors1982}
D.~Paget, Phys. Rev. B \textbf{25}, 4444, (1982).

\bibitem{MagneticFieldDependence}
P.L. Kuhns, A. Kleinhammes, T. Schmiedel, W.G. Moulton, P. Chabrier,
  S. Sloan, E. Hughes, and C.R. Bowers, Phys. Rev. B \textbf{55}, 7824 (1997).

\bibitem{Abragam}
A. Abragam, \textit{Principles of Nuclear Magnetism} (Oxford University Press, New York, 1996).

\bibitem{InhomoHyperfine}
C. Deng and X. Hu, Phys. Rev. B \textbf{72}, 165333 (2005).

\bibitem{SDR}
D. Paget, Phys. Rev. B \textbf{30}, 931 (1984).

\bibitem{Brunetti}
A. Brunetti, M. Vladimirova, D. Scalbert, H. Folliot, and A. Lecorre, Phys. Rev. B \textbf{73}, 121202 (2006).

%\bibitem{OPNMRold}
%T. Pietrass, A. Bifone, T. Room, and E.L. Hahn, Phys. Rev. B \textbf{53}, 4428%(1996).

\bibitem{OPsilicon}
A.S. Verhulst, I.G. Rau, Y. Yamamoto, and K.M. Itoh, Phys. Rev. B \textbf{71}, 235206 (2005).

\bibitem{CdTe}
I.J.H Leung and C.A. Michal,
Phys. Rev. B \textbf{70}, 035213 (2004).

\bibitem{OOQWs}
S. Pfalz, R. Winkler, T. Nowitzki, D. Reuter, A. Wieck, D. Hagele, and M. Oestreich,
Phys. Rev. B \textbf{71}, 165305 (2005).

%\bibitem{CdTe}
%I.J.H. Leung and C.A. Michal,
%Phys. Rev. B \textbf{70}, 035213 (2004).

%\bibitem{AKPthesis}
%A.K. Paravastu, Ph.D. Thesis. UC Berkeley (2004).

\bibitem{CarbonAcceptorPL}
K.D. Glinchuk, N.M. Litovchenko, A.V. Prokhorovich, and O.N. Stril'chuk, Semiconductors \textbf{35}, 384 (2001).

\end{thebibliography}

\end{document}